\documentclass[onecolumn,superscriptaddress,floatfix,preprintnumbers]{revtex4}
\usepackage{graphics,amssymb,amsmath,epsfig,color}
\usepackage{graphicx}

\begin{document}

\title{Lipid exchange promotes fusion of model protocells}


\author{Ziyan Fan}
\author{Yaam Deckel}
\author{Lauren A. Lowe}
\author{Daniel W.K. Loo}
\affiliation{School of Chemistry, Australian Centre for Astrobiology, ARC Centre of Excellence in Synthetic Biology, UNSW RNA Institute, UNSW Sydney, NSW 2052, Australia}

\author{Prof. Dr. Tetsuya Yomo}
\affiliation{Laboratory of Biology and Information Science, School of Life Sciences, East China Normal University, Shanghai, 200062, People’s Republic of China}

\author{Prof. Dr. Jack W. Szostak}
\author{Dr. Collin Nisler*}
\affiliation{Department of Chemistry, The University of Chicago, IL 60637, USA}

\author{Dr. Anna Wang*}
\affiliation{School of Chemistry, Australian Centre for Astrobiology, ARC Centre of Excellence for Synthetic Biology, UNSW RNA Institute, UNSW Sydney, NSW 2052, Australia}

\date{\today}
\maketitle

\section{Abstract}

Vesicle fusion is an important process underlying cell division, transport, and membrane trafficking. In phospholipid systems, a range of fusogens including divalent cations and depletants have been shown to induce adhesion, hemifusion, and then full content fusion between vesicles. This works shows that these fusogens do not perform the same function for fatty acid vesicles, which are used as model protocells (primitive cells). Even when fatty acid vesicles appear adhered or hemifused to each other, the intervening barriers between vesicles do not rupture. This difference is likely because fatty acids have a single aliphatic tail, and are more dynamic than their phospholipid counterparts. To address this, we postulate that fusion could instead occur under conditions, such as lipid exchange, that disrupt lipid packing. Using both experiments and molecular dynamics simulations, we verify that fusion in fatty acid systems can indeed be induced by lipid exchange. These results begin to probe how membrane biophysics could constrain the evolutionary dynamics of protocells.

\section{Introduction}
Lipid bilayer membrane fusion is a critical step underlying many processes essential to life, including cell division, membrane trafficking, and the transport of molecules~\cite{chernomordik_biomembrane_1987, chernomordik_membrane_2005, chernomordik_mechanics_2008}. For instance, the division of a cell requires the two lipid membranes to contact and then fuse in order to create two distinct cells from one mother cell. As a result, researchers have tried to understand how membrane fusion occurs and recapitulate the process in synthetic systems~\cite{chernomordik_mechanics_2008, terasawa_coupling_2012, litschel_freeze-thaw_2018}.

\begin{figure}
  \includegraphics[width=0.5\linewidth]{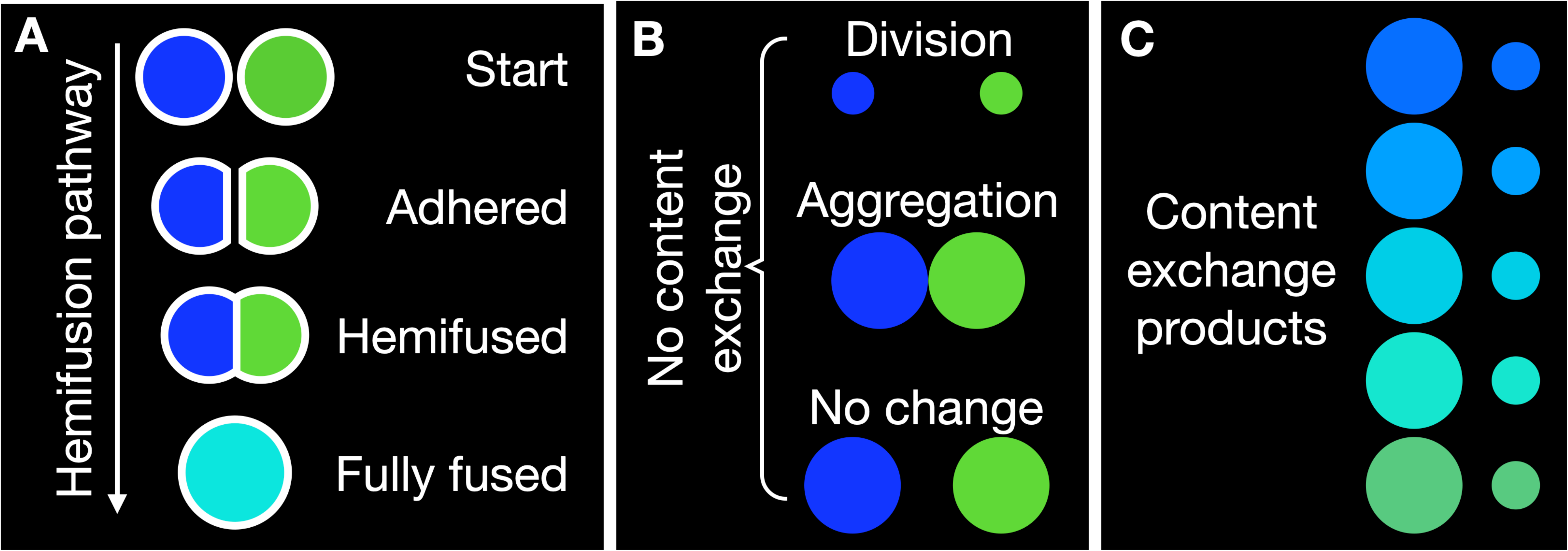}
  \caption{Schematics showing the different stages of vesicle fusion. (A) The hemifusion pathway for vesicle fusion includes first adhering the vesicles, then the vesicles sharing a hemifusion diaphragm, and finally, full fusion including content exchange. (B) Outcomes that do not involve full content fusion include vesicle division, aggregation, or no apparent change. (C) The presence of vesicles containing both dyes, resulting in intermediate colours, indicates content exchange from vesicles undergoing full fusion.}
  \label{fig:1}
\end{figure}

Membrane fusion is thought to proceed in a series of steps: the formation of a lipid stalk, the creation of a hemifusion diaphragm, the rupturing of this diaphragm via a fusion pore, then complete membrane fusion including content mixing (Fig.~\ref{fig:1} A). Exposing hydrophobic tails to aqueous media twice -- to achieve hemifusion, and then full fusion -- is fundamentally necessary to create a topologically distinct state. To minimize the energetic cost, it is thought that dehydration of the vicinity of the lipid is important~\cite{lentz_peg_2007}. 

It has also been suggested that \textit{lipid splay} is critical for membrane fusion~\cite{smeijers_detailed_2006, mirjanian_splaying_2010}. This is a process where diacyl lipids straddle the chasm between a merging pair of membranes thereby forming a hydrophobic bridge -- with one lipid tail inserted into one side and the other lipid tail inserted into the other. The two cells can then proceed to form the intermediate hemifused state, before achieving complete fusion~\cite{chernomordik_membrane_2005}.

There are many strategies for fusing membranes made from phospholipids. These include physicochemical strategies, such as using divalent cations (particularly Ca$^{2+}$)~\cite{warner_evolution_2012, ohki_phospholipid_1982} and depletion agents such as polyethylene glycol (PEG)~\cite{lentz_peg_2007}, and more biological ones such as soluble N-ethylmaleimide-sensitive factor attachment proteins receptor (SNARE) proteins~\cite{bock_snare_1999}, and cholesterol-tethered deoxyribonucleic acid (DNA)~\cite{chan_lipid-anchored_2008}.

In this work, we investigate the fusion of fatty acid vesicles. In general, there is little information on how membranes composed entirely of monoacyl lipids can fuse. Such molecules include fatty acids, lyso-lipids, and monoacyl glycerols, which are all intermediates of lipid synthesis~\cite{exterkate_growing_2018}. They are also intermediates of lipid digestion and are found in the intestinal tract~\cite{clulow_closer_2018}. However, monoacyl lipids by definition are incapable of lipid splay, leading us to question how fusion can be achieved in these systems.

Fatty acids are of particular interest because they are potential membrane components at the origins of life~\cite{wang_lipid_2019} and often used for making model primitive cells known as protocells. Fatty acids form membranes at a pH near their apparent p$K_\mathrm{a}$, and can support the insertion of small quantities of other lipids such as fatty alcohols, lyso-lipids, and monoacyl glycerols~\cite{wang_lipid_2019}. At the origins of life, the advent of compartmentalization in protocells and a physicochemical means of triggering compartment fusion would have been essential. Because content exchange between fatty acid vesicles has only been demonstrated via pH-induced phase changes, which also results in significant content loss~\cite{rubio-sanchez_thermally_2021}, we sought to find a route for content exchange without the vesicle assembly and disassembly necessitated by phase changes.

We use both experiments and simulations to probe fatty acid vesicle fusion. We find that fatty acid vesicles resist full fusion when using traditional fusion approaches i.e. use of divalent cations, depletants, increased salt concentration, and centrifugal forces. We also find that hemifused intermediates can persist for months and prevent content exchange for pure fatty acid systems. This is surprising given that hemifused intermediates only last 5-15 ns before complete fusion for phospholipid systems~\cite{marrink_mechanism_2003} and blended phospholipid/fatty acid systems~\cite{knecht_molecular_2007}. This time can be extended to the microsecond~\cite{kasson_control_2007} or minute~\cite{nikolaus_direct_2010} timescales, depending on the phospholipid composition~\cite{kasson_control_2007} and exact fusogenic conditions. Instead, we suggest that the potential disruption of lipid packing using lipid exchange may be a more efficient means of promoting fusion for fatty acid vesicles.

\section{Results and Discussions}

\subsection{Experimental attempts at fusion}

Two populations of giant unilamellar vesicles (GUVs) made from 5 mM oleic acid were made by a fatty acid self-assembly method~\cite{lowe_methods_2022, kindt_bulk_2020} as detailed in the Experimental section. In brief, each population was marked by an encapsulated dye and false-colored either blue or green (Fig.~\ref{fig:1} B). The vesicles were then diluted at least ten-fold to enhance contrast against the background for fluorescence imaging, then mixed under various conditions. Colors intermediate to blue and green, such as aqua or turquoise, would indicate content exchange between vesicles (Fig.~\ref{fig:1} C). 

\subsubsection{Attempted fusion with typical fusogens did not fuse vesicles}
Fatty acid membranes are negatively charged, which confers colloidal stability but also prevents them from adhering and then fusing. We tried a range of conditions (Table S1) to overcome the electrostatic repulsion between the membranes and encourage fusion.

\begin{figure}
  \includegraphics[width=0.5\linewidth]{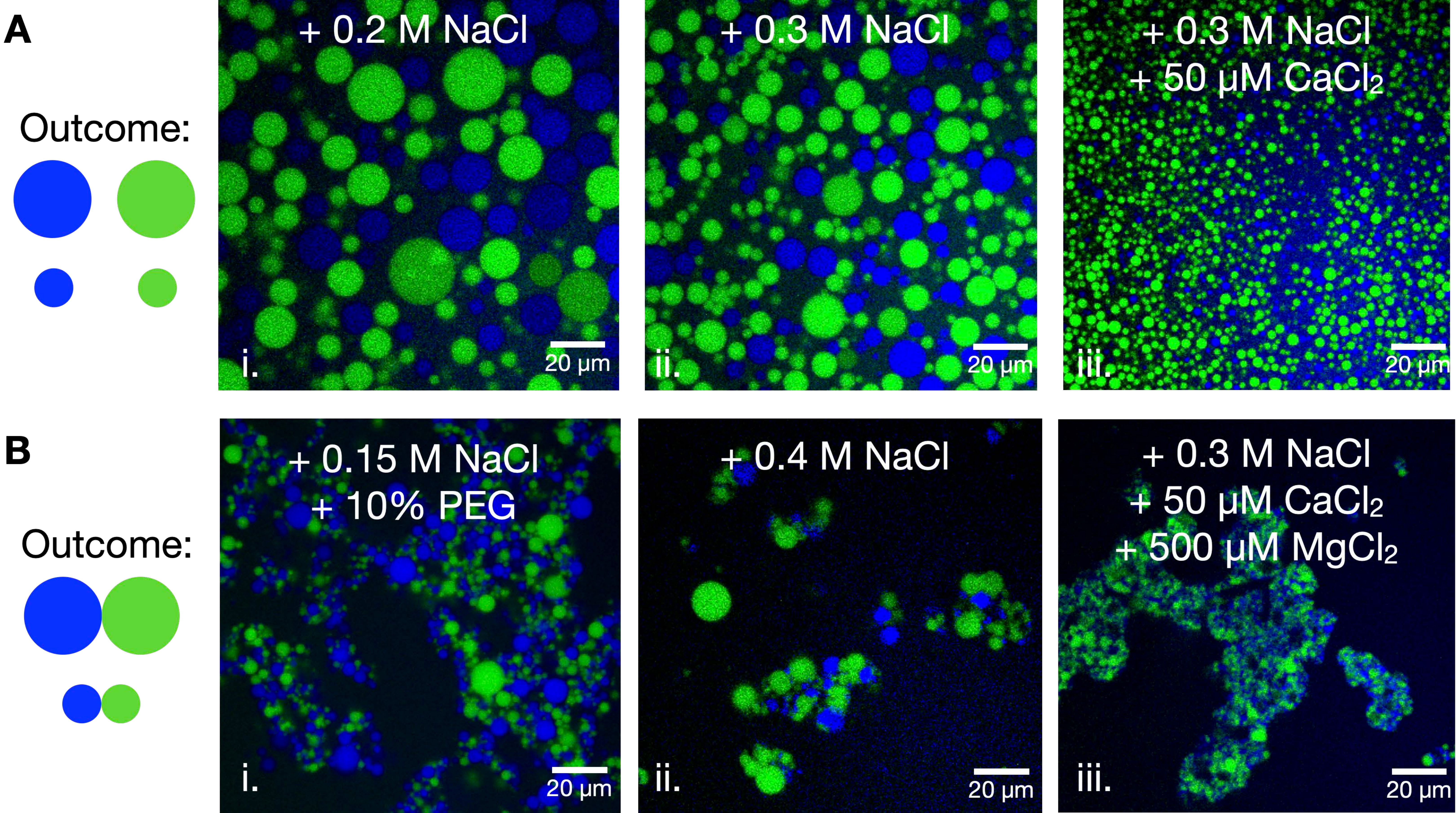}
  \caption{Oleic acid vesicles marked by either blue or green aqueous dyes were diluted, then mixed in the presence of various additives. (A) Some conditions resulted in vesicles not fusing or adhering including the addition of i. 0.2 M NaCl, ii. 0.3 M NaCl, and iii. 0.3 M NaCl and 50 µM CaCl$_2$. (B) Increasing the amount of fusogen present resulted in vesicles that adhered into large networks, but did not exchange contents or fully fuse. Conditions include the addition of i. 0.15 M NaCl with 10 \% PEG (w/v), ii. 0.4 M NaCl, and iii. 0.3 M NaCl and 50 µM CaCl$_2$ with 500 µM MgCl$_2$. For all the conditions explored see Table S1.}
  \label{fig:2}
\end{figure}

Under some conditions (Fig.~\ref{fig:2}A), vesicles appeared separate and non-adhering. In many cases, vesicles appeared smaller, suggesting potential vesicle division. This could have potentially occurred from the hyperosmotic shock drawing water out of the vesicles and increasing the excess surface area available to the vesicle. While the division of vesicles also requires the complete fusion of two lipid bilayer membranes, the conditions leading to division do not appear to result in vesicle fusion.

\begin{figure}
  \includegraphics[width=0.5\linewidth]{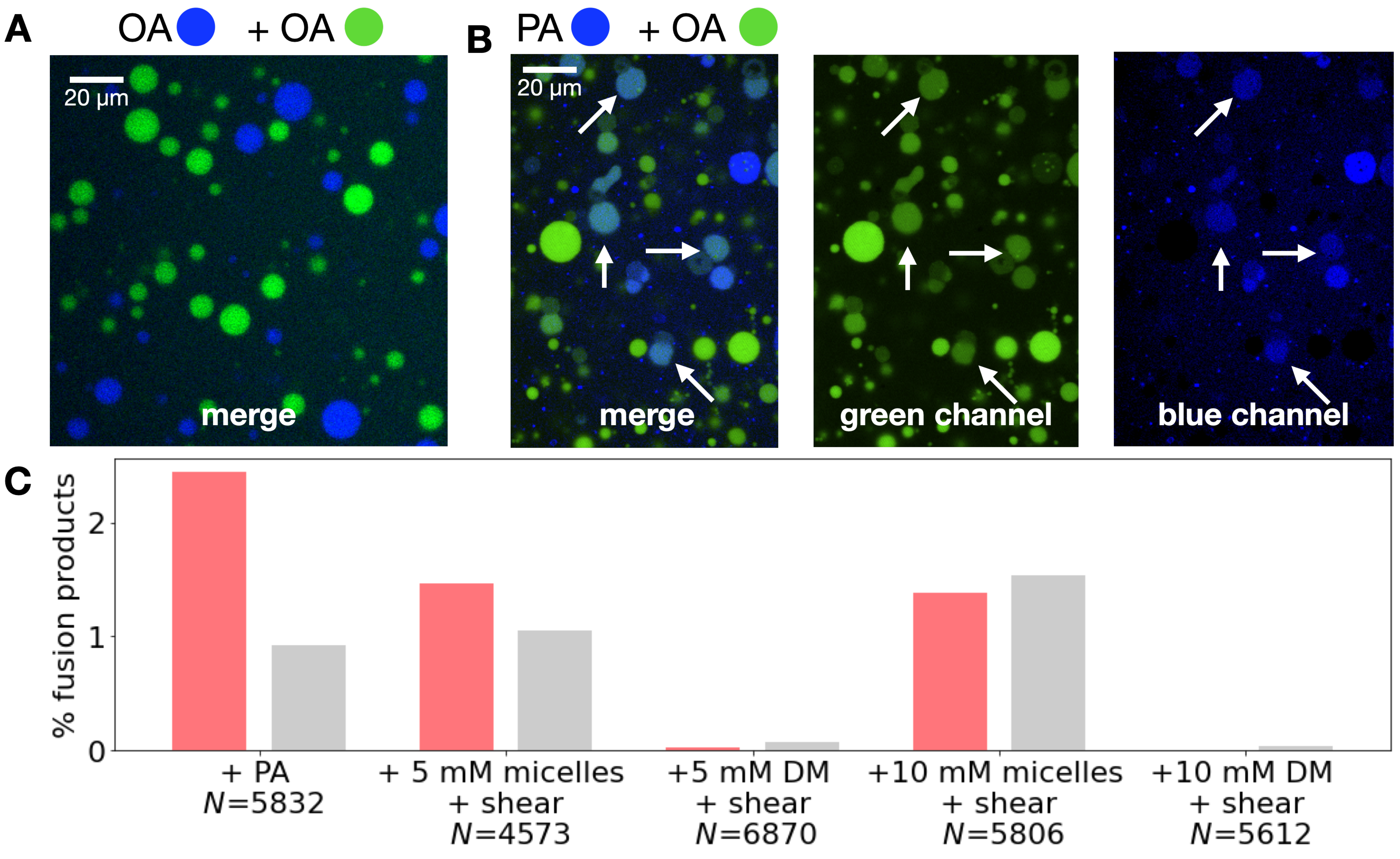}
  \caption{Only a few conditions were found to fuse fatty acid vesicles. (A) Mixing OA vesicles did not result in fusion, with the populations marked by blue and green dye remaining separate. (B) Adding a high concentration of PA vesicles (1 µL of 100 mM PA vesicles, in blue) near OA vesicles (green) did result in full fusion, with many vesicles containing colors intermediate to blue and green similar to that shown in Fig.~\ref{fig:1}C (white arrows). (C) Controlled measurements of mixing equal amounts of OA GUVs with PA GUVs, or OA vesicles with micelles followed by shear (24 hours of orbital shaking), led to full fusion and content exchange between vesicles. Without the 24 hours of shaking, only the addition of PA vesicles was found to induce fusion. The total numbers of vesicles counted across two replicates per condition are indicated. DM = `dummy' micelles.}
  \label{fig:3}
\end{figure}

To increase the vesicle attractive interaction potential to encourage fusion, we added a depletant (poly(ethylene glycol) PEG), monovalent salt (NaCl), and divalent salts (MgCl$_2$, CaCl$_2$, CoCl$_2$). While the vesicles did adhere to each other (Fig.~\ref{fig:2}B), there was no fusion between the adhered compartments. Indeed, under all of the conditions tested (Table~S1), the two populations of oleic acid vesicles with different colors remained separate (as in Fig.~\ref{fig:3} A). None of the typical physicochemical fusogens (Ca$^{2+}$, Mg$^{2+}$, PEG, centrifugation) were found to fuse these membranes.

Vesicle doublets, which are potential intermediates in the fusion process, were consistently observed in samples made of oleic acid including months after sample preparation. To develop a deeper understanding of the resistance to fusion, we analyzed these doublets and confirmed that they were hemifused and not simply adhered (see SI, Fig. S1-S2)~\cite{warner_evolution_2012, warner_hemifused_nodate}. Rather than the typical timescales of nanoseconds to seconds~\cite{nikolaus_direct_2010}, the hemifused `intermediates’ in fatty acid systems were observed to be stable over hours of observation and persisted in samples for months. Such stability could prevent full fusion from occurring. We therefore sought a different strategy to the conventional methods for phospholipid vesicles, to achieve full fusion.

\subsubsection{Lipid exchange can be used to trigger fusion}

Because the exposure of the lipids' hydrophobic tails to aqueous media is necessary to fuse two membranes, we sought to determine whether disturbances in lipid packing could lead to membrane fusion. To achieve a disturbance in lipid packing, we leveraged the fact that monoacyl lipids have a larger soluble monomer fraction relative to phospholipids~\cite{wang_lipid_2019}: When additional fatty acids are added (for example, from a solution of lipid-rich ethanol) to a solution of vesicles, some fatty acids aggregate with nearby fatty acids (\textit{de novo} vesicle formation), whereas others will partition onto the pre-existing membranes~\cite{kindt_bulk_2020, chen_kinetic_2004, peterlin_growth_2009}. Thus instead of using dehydration to minimize the energetic cost of exposing lipid tails to water, the disturbance in lipid packing could be driven by the rapid addition of lipids onto the membrane.



\textbf{Condition 1 -- mixing oleic acid vesicles with palmitoleic acid vesicles} \\

It is known that mixing vesicles made from fatty acids of different chain lengths results in rapid lipid exchange between the vesicles~\cite{toparlak_population-level_2021}, potentially providing disturbances to lipid bilayer packing that could aid fusion. We therefore created vesicles from oleic acid (OA; 18 carbons, false-colored green), and vesicles from palmitoleic acid (PA; 16 carbons, false-colored blue) to determine the effects of mixing vesicles made of different lipids.

We found that depositing 1 µL of 100 mM PA vesicles near 0.5 mM OA vesicles yielded large numbers of fusion products (Fig.~\ref{fig:3} B). To quantify the effect, we created two samples of GUVs, diluted them forty-fold to provide contrast between the internal vesicle contents and the background, and mixed the two vesicle populations in equal proportion. We found that approximately 1--2 \% of vesicles ($N$ = 5832) were fusion products, after letting the sample equilibrate for 30 mins (Fig.~\ref{fig:3} C). 

\textbf{Condition 2 -- mixing oleic acid GUVs with oleate micelles} \\

The packing of lipid bilayers can also be disturbed by the addition of fatty acid micelles. This procedure has been shown to lead to the incorporation of fatty acids onto the membrane, leading to vesicle shape changes including fission~\cite{kindt_bulk_2020}. To determine the effect of micelle addition on fusion, we created two populations of OA GUVs labelled with either xylenol orange (false-colored green) or pyranine (false-colored blue), diluted them forty-fold to a final concentration of 0.5 mM oleic acid, and mixed them together. After the addition of either 5 or 10 equivalents of micelles to the mixture, we found no fusion after 24 hours. We also attempted agitating the samples on an orbital shaker for 1 hour after micelle addition, and found that there was no fusion.

However, after orbital shaking for 24 hours following micelle addition, we found substantially higher levels of fusion. For 5 equivalents ($N$ = 4573, final pH 8.01) and 10 equivalents ($N$ = 5806, final pH 8.11) of micelles added, approximately 1 \% of vesicles were fusion products (Fig.~\ref{fig:3} C).

We hypothesize that the shaking enables further lipid exchange and opportunities for vesicle-vesicle collisions. To test this, we created two populations of vesicles labeled with membrane-bound dyes TopFluor PC (green) and Lissamine Rhodamine PE (red). Because the lipid dyes can not freely exchange owing to their membrane-anchoring phospholipid moieties, we were surprised to find a substantial amount of lipid membrane mixing following overnight orbital shaking (Fig.~S3), suggesting that the lipid membranes are exchanging via some mechanism. We have considered the mechanism for the exchange of the dye-labelled phospholipids. Oleate micelles may be able to remove these lipids and shuttle them between vesicles. Because there is no dyed phospholipid exchange without shear, and shear should not affect whether micelles form or not, we think it is more likely that transient hemifusion between collided vesicles is the main contributor to phospholipid exchange. However, despite the lipid exchange, no content exchange was found to occur (in the absence of micelles) even after overnight agitation (Fig.~S3), suggesting that collision-based lipid transfer alone is not sufficient to engender fusion.

\textbf{Condition 3 -- mixing oleic acid GUVs with `dummy' micelles} \\
We have previously shown that increasing the pH of fatty acid vesicles can lead to a decrease in lipid packing order~\cite{lowe_subtle_2022}. To determine whether it was an increase in pH from micelle addition that caused the low levels of fusion found under Condition 2, we decided to repeat Condition 2 but with an increase in pH from `dummy' micelles (dilute NaOH) rather than micelle addition. 

Again, we found that there was no fusion after 1 hour even with a pH increase from adding `dummy' micelles. Again, we placed the mixed sample onto an orbital shaker for 24 hours. For 5 mM `dummy' micelles added, 0.06 \% of vesicles (4 out of $N$ = 6870, final pH 8.04) were fusion products and for 10 mM `dummy' micelles added, 0.02 \% of vesicles (1 out of $N$ = 5612, final pH 8.16) were fusion products (Fig.~\ref{fig:3} C). This supports the notion that a slight increase in pH and agitation may not be enough to engender fatty acid vesicle fusion. Instead, lipid exchange on the membrane is necessary.

Taken together, our experimental results show that the divalent cations and/or depletants (PEG) used for phospholipid membrane fusion do not perform the same function for fatty acids. One possible reason is that single-chained amphiphiles, being much more dynamic than their double-chained counterparts, are able to rapidly relieve mechanical and bending stresses that would otherwise be relieved by fusion~\cite{bruckner_flip-flop-induced_2009}. Alternatively, carboxylic acid lipid head groups could be more difficult to dehydrate than phospholipid head groups. Another possible reason is that the absence of lipid splay owing to the presence of a single lipid tail is preventing the initiation of fusion.

\subsection{MD simulations of fusion between fatty acid vesicles}

\begin{figure}
  \includegraphics[width=\linewidth]{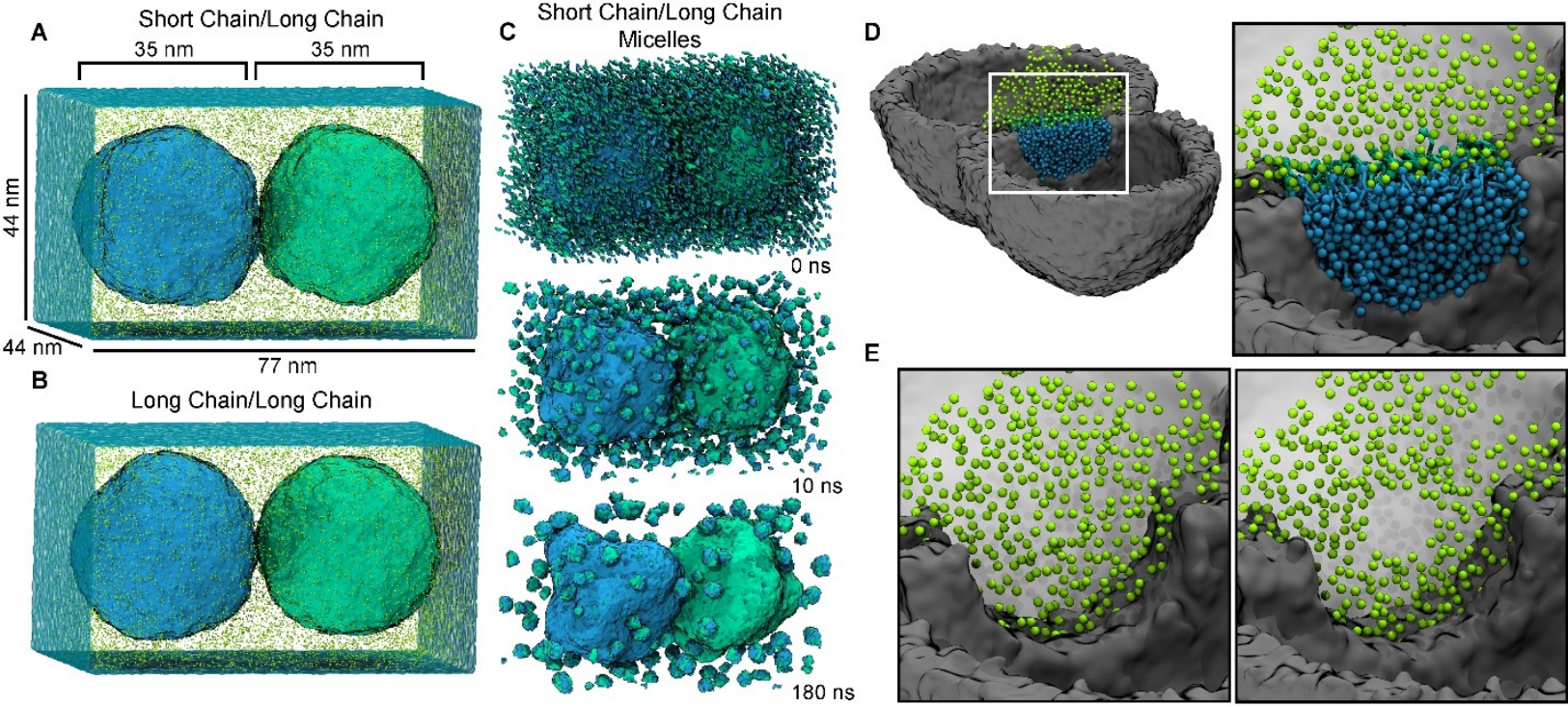}
  \caption{Overview of simulation systems. (A) Sys1: Molecular image of the short-chained/long-chained system, with the short-chained vesicle in blue and long-chained vesicle in green. Water is shown in transparent surface representation, with ions shown as yellow spheres. Dimensions of the system and vesicles are indicated. (B) Sys2: Molecular image of the long-chained/long-chained system, with vesicle 1 in blue and vesicle 2 in green. Dimensions of the system are the same as in (A). (C) Sys3: Molecular image of the short-chained/long-chained system with micelles at different points during equilibration. Water and ions are excluded for clarity. Colors correspond to those in (A). (D) (Left) Molecular image of the short-chained/long-chained system, where the surface of each vesicle is shown as a grey surface sliced halfway through the system through the \textit{x-y} plane. (Right) A zoomed-in view showing interfacial molecules from vesicle 1 (blue) and vesicle 2 (green; obscured), and showing the layer of water and solvent molecules as spheres. (E) An identical view as in the zoomed in view of (D) without the interfacial molecules, showing before (left) and after (right) the application of an outwardly-directed radially symmetric force to create a transient vacuum between vesicles.}
  \label{fig:MD1}
\end{figure}

To gain mechanistic insight into our experimental results, we conducted a series of coarse-grained (GG) molecular dynamics (MD) simulations to determine how chain length and the presence of micelles affect the interaction dynamics of vesicles. For this purpose, three CG systems were built. The first consisted of a long-chained vesicle and a short-chained vesicle, both 17.7 nm in radius, placed 360 Å apart from their centre of mass and with no free fatty acids (Sys1 short/long; for details on system construction see Methods). The second consisted of two long-chained vesicles, both 17.7 nm in radius, placed 360 Å apart from their centre of mass, also with no free fatty acids (Sys2 long/long). The third was identical to the first but with 4,620 short-chained and 4,620 long-chained fatty acids added randomly to the surrounding medium (Sys3 short/long/micelles; Fig.~\ref{fig:MD1} A-C). The concentration of free fatty acids was chosen to be high enough such that small micelles would readily form upon equilibration, but low enough to avoid formation of bilayer structures.

\subsubsection{Shorter chain length and presence of micelles enhances the association between vesicles at equilibrium}

Because the initiation of fusion relies on stochastic fluctuations in the membrane surface, formation of a larger interface between opposing vesicles would likely result in a higher probability of a fusion event. Both Sys1 (short/long) and Sys3 (short/long/micelles) exhibited an adhesive interface after 500 ns of equilibration (Sim1 and Sim3), but the presence of free fatty acids greatly enhanced the interaction. Within 1 ns, the free fatty acids coalesced to form small ($\sim$30 Å diameter) micelles (Fig.~\ref{fig:MD1} C), which fused with the vesicles and with each other to form larger micelles ($\sim$50 Å diameter) after $\sim$180 ns (Fig.~\ref{fig:MD1} D).

\begin{figure}
  \includegraphics[width=\linewidth]{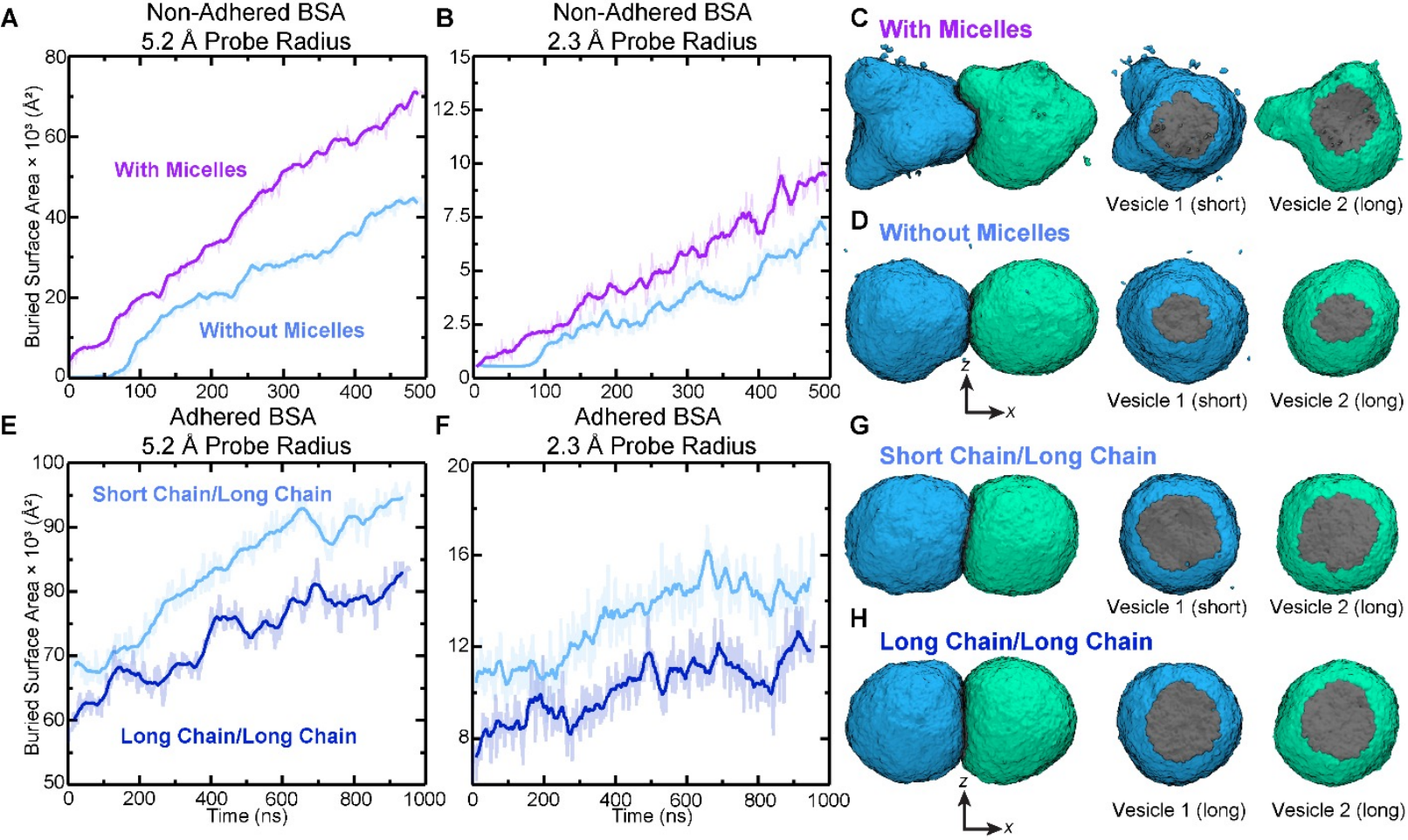}
  \caption{Buried surface area (BSA) between adhered vesicles during the equilibrium simulation. (A-B) Buried surface area of the short-chained/long-chained vesicle system with (Sys3) and without (Sys1) micelles over 500 ns of equilibrium simulation using a 5.2 Å probe radius (A), corresponding to interfacial area including the monolayer of ions and water beads, and a 2.3 Å probe radius (B), corresponding only to surface area that is a direct contact between vesicles. In each case, the presence of micelles resulted in an increase in buried surface area. (C-D) Molecular images showing the final structure of the vesicles in surface representation, with interfacial area shown in grey. (E-F) Buried surface area of the short-chained/long-chained vesicle system (Sys1) and long-chained/long-chained vesicle system (Sys2) over 956 ns of adhered equilibrium simulation using a 5.2 Å probe radius (A) and a 2.3 Å probe radius (B). The greater deformability of the short-chained vesicle resulted in an increase in buried surface area. (G-H) Molecular images showing the final structure of the vesicles in surface representation, with interfacial area shown in grey. In (A-B) and (E-F), transparent traces are raw data and solid lines are 12 ns running averages.}
  \label{fig:MD2}
\end{figure}

We used buried surface area (BSA) as a metric of association between vesicles and found this was enhanced for the presence of the shorter-chain length vesicle, and the presence of micelles. Using a probe radius of 5.2 Å, which corresponds to twice the radius of a water bead and thus measures the entire interfacial area between both vesicles (see Methods), Sys1 (short/long) exhibited a buried surface area of 43,555 Å$^2$, while the buried surface area in Sys3 (short/long/micelles) was much larger at 69,621 Å$^2$ (Fig.~\ref{fig:MD2} A, C-D). Using a probe radius of 2.3 Å, which measures only direct contact between vesicles, the buried surface area of Sys1 (short/long) was 6,430 Å$^2$ while that of Sys3 (short/long/micelles) was again larger 9,405 Å$^2$ (Fig.~\ref{fig:MD2} B). In Sys3 (short/long/micelles), the fusion of micelles with the vesicles resulted in excess surface area, inducing significant fluctuations in the vesicle surfaces (Fig.~\ref{fig:MD2}). This disruption of surface equilibrium, combined with a presumed depletion effect from the presence of the micelles, is likely responsible for the significant increase in buried surface area relative to the system without micelles, and provides a molecular explanation for the presence of fusion products between OA vesicles only in the system with micelles present (experimental condition 2).

Next, steered molecular dynamics (SMD) was used to induce a more extended interface between vesicles in Sys1 (short/long) and  Sys2 (long/long; Sim4 and Sim5). After 50 ns of equilibration, a constant velocity force was applied to the left-most vesicle in each system (vesicle 1 of Fig.~\ref{fig:MD2} G-H) at a speed of 0.1 nm/ns for $\sim$125 ns. The force was then released, and each system was allowed to equilibrate for 956 ns (Sim6 and Sim7). In both systems, a stable interface consisting of a single layer of solvent and Na$^+$ ions facilitated adhesion between vesicles (Fig.~\ref{fig:MD1} D). After equilibration, the buried surface area of Sys1 (short/long) was 95,572 Å$^2$ while that of Sys2 (long/long) was 83,264 Å$^2$ using a probe radius of 5.2 Å (Fig~\ref{fig:MD2} E). Using a probe radius of 2.3 Å, the buried surface area was 14,997 Å$^2$ in Sys1 (long/long) and 13,172 Å$^2$ in Sys2 (short/long) (Fig.~\ref{fig:MD2} F). This increase in buried surface area in Sys1 (short/long) is likely a result of the higher deformability of the shorter-chained vesicle, which formed a concave interface with the opposing vesicle, allowing a greater overlap between vesicles (Fig.~\ref{fig:MD2} G-H). The deformation may also be a result of the interface being asymmetric. In Sys2 (long/long), the higher membrane rigidity and symmetry of the system resulted in a flatter interface and reduced buried surface area. Overall, the shorter-chained vesicle formed a more extended interface with an opposing vesicle, increasing the surface area available for the initiation of a fusion event, and this effect is significantly enhanced by the presence of micelles in the surrounding medium.

\subsubsection{Chain length alters the molecular composition of the adhesive interface between vesicles}

One factor that could affect the probability of initiation of a fusion event is the molecular composition of the interface between vesicles. Analysis of the buried surface area using the smaller 2.3 Å probe radius (Fig.~\ref{fig:MD2} F) suggested that the interface formed by a short-chained vesicle presents a greater opportunity for direct contact between vesicles, but this result does not account for the charge state of the molecules involved. Previous work on mixed phospholipid/fatty acid systems indicates that a lower pH enhances fusion due to the presence of fewer negatively charged headgroups~\cite{zellmer_temperature-_1994}, while at higher pHs fatty acids generate more positive curvatures~\cite{lahdesmaki_membrane_2010}, which would hinder fusion. As such the ratio of neutral to negatively charged molecules in the interface was quantified for Sys1 (short/long) and Sys2 (long/long) during the 956 ns equilibration. 

\begin{figure}
  \includegraphics[width=\linewidth]{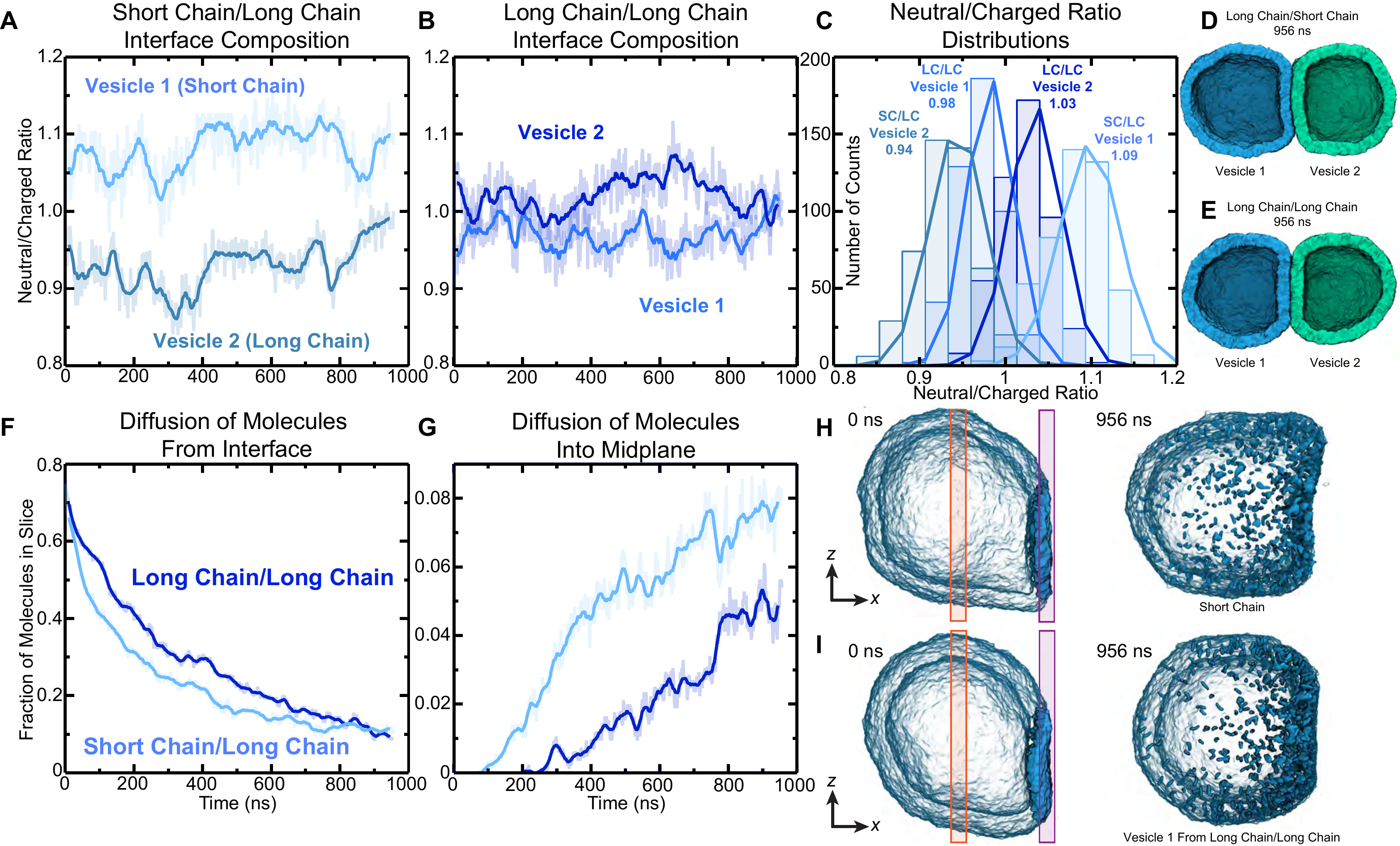}
  \caption{Interface identity and dynamics differs between long-chained and short-chained vesicles. The ratio of neutral molecules to negatively charged molecules in the interface is shown for each vesicle in the short-chained/long-chained system (A) and long-chained/long-chained system (B). The short-chained vesicle exhibits fewer negatively charged molecules in the interface over the 956 ns simulation. (C) Histograms of the neutral/charged ratio in the interface for all vesicles with fits to a normal distribution. The mean for each fit is shown in the respective colors. (D-E) Molecular images showing the final structure of the vesicles in surface representation sliced halfway through each vesicle. In (D), the short-chained vesicle is shown in blue, and the long-chained vesicle in green. (F) Shown is the fraction of molecules found in the first \textit{y-z} volume slice (magenta box in H and I) that originated from the interface at the start of the simulation, representing the diffusion of molecules away from the interface during the simulation. (G) Shown is the fraction of molecules found in a \textit{y-z} volume slice at the midplane of the vesicle that originated from the interface at the start of the simulation, representing the diffusion of molecules into this region of the vesicle during the simulation. Molecular images for the short-chained/long-chained system (H) and long-chained/long-chained system (I). Interfacial molecules are shown at the start of the simulation (0 ns) in opaque surface representation with the rest of the vesicle in transparent surface, and the same residues shown at the end of the simulation (956 ns). The volume slice from (F) is shown in a pink box, and the volume slice from (G) is shown in an orange box. In (A-B) and (F-G), transparent traces are raw data and solid lines are 12 ns running averages.}
  \label{fig:MD3}
\end{figure}

Briefly, at each frame of the simulation, the interface molecules from vesicle 1 were defined as those that are within 21 Å of any molecule from vesicle 2 and vice-versa, and the number of neutral versus negatively charged species was counted from this set for each vesicle. Over the course of the equilibration, all long-chained vesicles (both vesicles from Sys2 (long/long), and vesicle 2 from Sys1 (short/long)) exhibited a neutral/charged ratio $\sim$1 or below (Fig.~\ref{fig:MD3} A-B). In contrast, the neutral/charged ratio for the short-chained vesicle remained close to $\sim$1.1, suggesting the interface formed in Sys1 consists of fewer charged molecules. This is reminiscent of lipid sorting, as seen in mixed phospholipid systems~\cite{kawamoto_coarse-grained_2015}, and could also be playing a role in generating the preferred curvatures in our system. Uncharged headgroups occupy less area than charged headgroups, and could help favor the negative curvature required for stalk formation. This is supported by the asymmetry induced in the interface in Sys1 but not Sys2 where the difference in the ratio of neutral/charged molecules between interfacing vesicles is far greater in Sys1 ($\sim$0.15) compared to Sys2 ($\sim$0.05), and by visually observing the negative curvature induced in the short-chained vesicle (Fig.~\ref{fig:MD3} C-E).

As vesicles are brought into proximity, a buildup of negative charge at the interface from deprotonated head groups could be relieved by diffusion. Therefore, one factor that could lead to a difference in interfacial identity is an increase in dynamics afforded by the shorter-chained fatty acids. To test this, the diffusion of molecules from the interface was quantified for Sys1 (short/long) and Sys2 (long/long) during the 956 ns equilibration. For vesicle 1 of each system, molecules in the interface (defined as above) in the first frame of the simulation were tracked throughout the equilibration. The vesicle was divided into a series of 12 transversal volume slices in the \textit{y-z} plane (Fig.~\ref{fig:MD3} H-I), and the fraction of molecules contained in that volume slice that originated from the interface was quantified. In the first volume slice, which corresponds to the interface in each system, both systems exhibited an immediate reduction in original molecules (Fig.~\ref{fig:MD3} F, H-I). However, the dynamics of this diffusion for Sys1 (short/long) were faster, leading to a more rapid turnover of molecules in the interface for the short-chained vesicle. Diffusion of original interfacial molecules into the volume slice taken at the midpoint of the vesicle was also more rapid and resulted in more molecules in this region of the vesicle for Sys1 (short/long; Fig.~\ref{fig:MD3} G, H-I). 

These results suggest that faster diffusion in the shorter-chained vesicle leads to more turnover in interface molecules, potentially enabling the decreased negative charge in the interface leading to less electrostatic repulsion, and more favorable curvature for stalk formation. This mechanism is likely generalizable to adhered membrane systems that exhibit a disparity in dynamics, where differential rates of diffusion or flip-flop of membrane components in each membrane drive an asymmetry in charge or composition at the interface.

Other methods can also be used to manipulate charge and curvature. For reducing electrostatic repulsion, a transient decrease in pH could help adhesion. However, care must be taken, because fatty acid vesicles collapse into neat oil droplets below the narrow pH range of vesicle formation~\cite{kindt_bulk_2020}. The pHs explored in this present work are already at the lower end of fatty acid vesicle formation. For generating curvature, uncharged headgroups occupying less area than charged headgroups could help favor the negative membrane curvature required for stalk formation. Evidence for this phenomenon comes from both simulations and experiments, where membranes composed of phosphatidylethanolamine (PE) head group lipids exhibit significantly more negative curvature than those composed of the larger phosphatidylcholine (PC) head group lipids~\cite{kozlovsky_lipid_2002, chernomordik_shape_1985}. Whilst fatty acids can become uncharged and protonated at lower pHs, thus having a reduced headgroup size, the self-assembly of membranes is compromised if the fraction of uncharged fatty acids is too low, as discussed above. The presence of lipids with smaller headgroups, such as fatty alcohols, could also potentially help in generating the required negative curvature for stalk formation.

\subsubsection{Induced formation of intermembrane stalk results in pore formation only in the short-chained vesicle system}

\begin{figure}
  \includegraphics[width=0.7\linewidth]{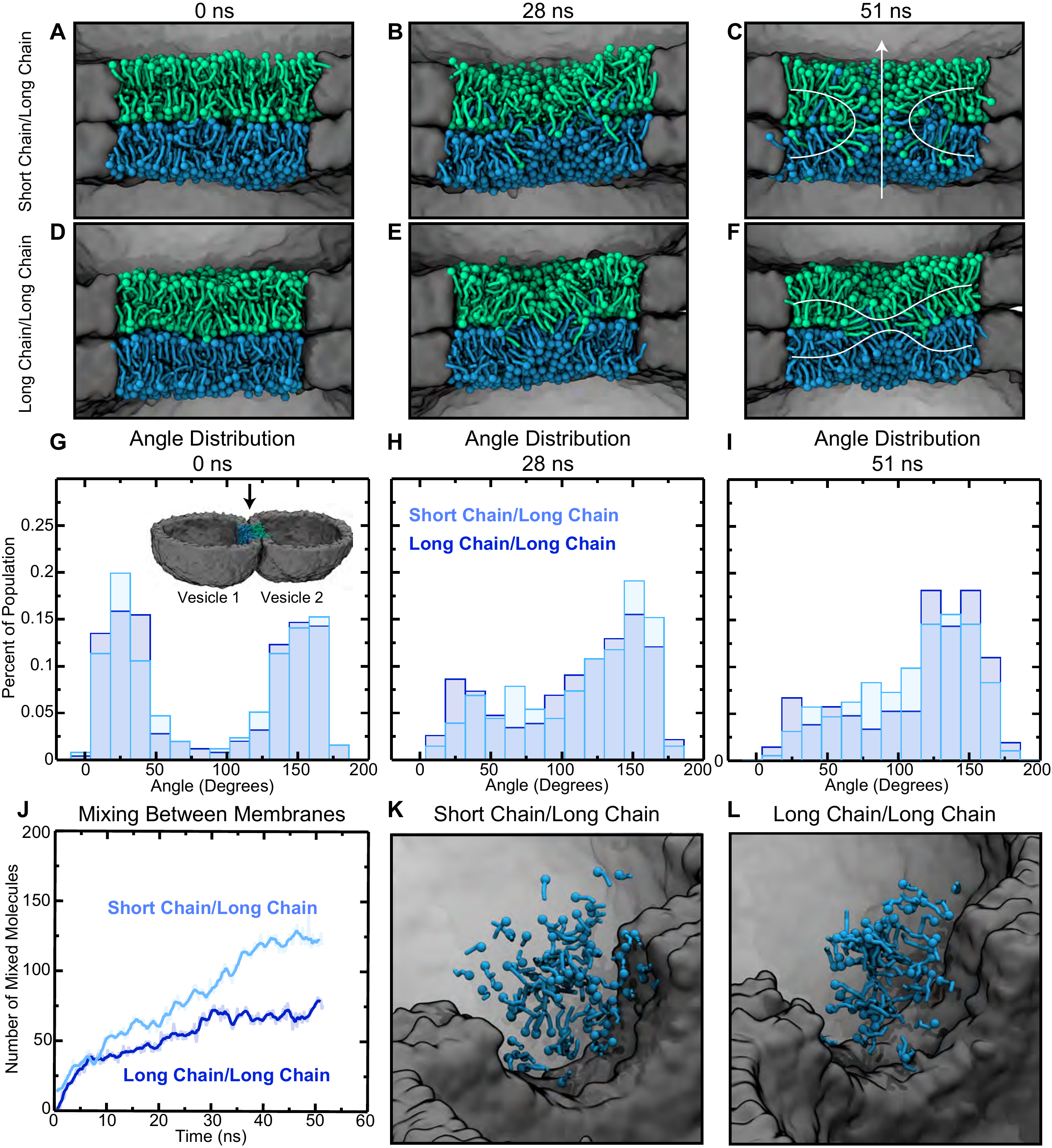}
  \caption{Interfacial dynamics during induced fusion between vesicles. (A-C) The short-chained/long-chained system during the fusion simulation at timepoints indicated as viewed from the inset in (G), where a pore and continuous membrane between vesicles formed between opposed vesicles within 50 ns. In (C), the midplane is shown in white and the pore is indicated by the white arrow. (D-F) The long-chained/long-chained system during the fusion simulation. Over the same timescale as the short-chained/long-chained system, this system did not form a pore and instead remains stable in the hemifused state. In (F), the midplane is shown in white. (G-I) Distribution of angles formed between molecules in the interface from vesicle 1 and the x-axis at the timepoint indicated. The long-chained molecules exhibit a greater retention of the initial bimodal distribution, corresponding to fewer disordered molecules. (J) Shown are number of molecules originating from vesicle 1 that are in contact with at least 4 molecules originating from the opposing vesicle, representing the degree of mixing between membranes. Molecular images of the short-chained/long-chained (K) and long-chained/long-chained (L) systems at 50 ns with surrounding vesicle surface shown in grey, and molecules from vesicle 1 that are in contact with at least 4 molecules from the opposing vesicle shown in blue licorice representation. The short-chained vesicle exhibits a higher number of mixed molecules in agreement with (J), and a wider spatial distribution at the interface. In (J), transparent traces are raw data and solid lines are 6 ns running averages.}
  \label{fig:MD4}
\end{figure}

Because the adhesive interface between vesicles was stable over the 956 ns of equilibration (Fig.~\ref{fig:MD1} D), and stochastic events that would lead to the formation of a stalk are not accessible on the timescales presented here, formation of a stalk was induced between vesicles in Sys1 (short/long) and Sys2 (long/long) to study the formation of a fusion pore in these systems (Sim8 and Sim9). An outwardly-directed radially symmetric force was applied to interfacial ions and solvent molecules to briefly create a vacuum between membranes. This force was maintained such that only solvent and ions that are within 20 Å of the center of the interface experience the force that keep them outside of this region of the interface (Fig.~\ref{fig:MD1} E). Within several nanoseconds, the presence of a transient vacuum drove the formation of a stalk between opposed membranes in both systems, and after $\sim$10 ns both systems exhibited a hemifusion diaphragm (Fig.~\ref{fig:MD4} A-B and D-E). While the exact transition from a stalk to a hemifusion diaphragm is difficult to identify in such a dynamic system, a hemifused state was identified by the presence of a continuous outer leaflet between opposed vesicles, and the mixing of tail groups between inner-leaflet molecules (Fig.~S4). The presence of a vacuum between vesicles is highly unfavorable and as such the formation of the hemifusion diaphragm is rapid in both systems, with the dynamics of this transition being slightly faster in Sys1 (short/long) (Fig. S4).

At $\sim$50 ns, this hemifused state was destabilized in Sys1 (short/long), and a fusion pore formed between vesicles, with each vesicle sharing a fully continuous membrane (Fig.~\ref{fig:MD4} C). However, over the same timescale, Sys2 did not form a fusion pore, and instead remained in a stable hemifused state (Fig.~\ref{fig:MD4} F). This suggests that formation of a localized hemifused state, resulting from a transient fluctuation such as packing defects or protrusions, is more likely to lead to a fusion pore with shorter-chained fatty acids. The formation of hemifused structures without fusion pores can also help to explain our experimental results, which reveals the presence of lipid mixing between two vesicle populations upon shear, but not content exchange (Fig.~S3).

To attempt to identify the molecular mechanism that results in pore formation in Sys1 (short/long) but not Sys2 (long/long), the disorder of molecules from vesicle 1 in each system was quantified. First, for every lipid molecule at the interface that falls within a 35 Å circle centered at the same location as the application of force on interfacial solvent and ions, a vector was defined between the last and first bead of each molecule. This vector was then projected onto the \textit{x-y} plane, and the angle it made with the \textit{x}-axis (from 0-180$^\circ$) was measured during the 51 ns fusion simulation. Membranes that exhibit more disorder will populate the region between 0-180$^\circ$, while less disorder will result in a more bimodal distribution that tends to either 0$^\circ$ or 180$^\circ$. At the start of the simulation, both Sys1 (short/long) and Sys2  (long/long) exhibited a bimodal population, as expected of a stable membrane (Fig.~\ref{fig:MD4} G). As the simulation proceeded, both systems lost this bimodal feature to some extent, but the membrane composed of longer-chained fatty acids maintained its population of molecules orthogonal to the membrane interface ($\sim$0-50$^\circ$) more readily than the shorter-chained fatty acids (Fig.~\ref{fig:MD4} H), indicating the retention of a more stabilized membrane packing. By the end of the simulation, Sys1 (short/long) lost the bimodal characteristic completely, while this feature was retained in Sys2 (long/long; Fig.~\ref{fig:MD4} I). This enhanced packing disorder in the shorter-chain membrane or retention of a more ordered state in the longer-chain membrane provides a molecular explanation for the stabilized hemifused state in Sys2 (long/long).

Finally, the degree of mixing between opposing membranes was characterized to determine how readily vesicles could exchange membrane components in these two systems during fusion. For Sys1 (short/long) and Sys2 (long/long), every molecule from vesicle 1 that was in contact with at least 4 molecules from vesicle 2 was counted, which was taken as a measure of mixing between membranes. Throughout the entire 56 ns fusion simulation, Sys1 (short/long) exhibited a higher degree of mixing compared to Sys2 (long/long; Fig.~\ref{fig:MD4} J). Additionally, by the end of the simulation, the mixing exhibits a broader spatial distribution in Sys1 (short/long) compared to Sys2 (long/long; Fig.~\ref{fig:MD4} K-L). Overall, these results suggest that enhanced disorder in the shorter-chain vesicle destabilizes the hemifused state and leads to a fusion pore with a more homogenized interface due to enhanced mixing between membranes.

\section{Discussion}

Here we discuss our experimental and molecular dynamics simulation results in the context of full fusion and content exchange between fatty acid vesicles requiring two steps, as for phospholipids~\cite{chernomordik_membrane_2005}. The first step is the exposure of the lipid's hydrophobic tails to aqueous media to create a fusion stalk. The second step involves the creation of a fusion pore to destabilize a hemifusion diaphragm, thereby enabling content exchange between two vesicles.



\subsection{Initiating fusion}
\subsubsection{Role of lipid splay}

Despite the extensive interaction both in equilibrium and upon a more extended interface through the application of steered molecular dynamics, none of our simulated systems exhibited spontaneous fusion or formation of a stalk-like structure between vesicles, as has been observed in previous simulations involving phospholipids~\cite{smeijers_detailed_2006, marrink_mechanism_2003, knecht_molecular_2007}. While many of the models used in those studies correspond to a vesicle radius of $\sim$11-17 nm and thus would experience a higher bending energy and propensity for fusion than for our system, another likely explanation for this discrepancy is the role of phospholipids in the initiation of fusion. Lipid splay was identified to be essential in this process in multiple simulation studies~\cite{smeijers_detailed_2006, mirjanian_splaying_2010, stevens_insights_2003}, a structural feature not available to pure fatty-acid systems. 

Thus, the initiation of a stalk in fatty acid membranes may depend on other stochastic events that were not observed on the simulation timescale sampled in our simulations. Disturbances in lipid packing during rapid lipid exchange on the membrane could lead to transient fusion intermediate structures that are lower in energy. Indeed, it was theoretically shown that even small out-of-plane movements of lipids, including tilting or protruding from the monolayers, could dramatically decrease the energies needed to form fusion intermediates~\cite{kuzmin_quantitative_2001}. This is consistent with our experimental results, which suggest that lipid exchange between vesicles made from two different fatty acids directly boosts the probability of fusion. 

Further evidence for this mechanism is observed in the simulations, where the shorter chain fatty acid molecules readily departed the vesicle bilayer and either inserted into the original vesicle, or exchanged with the neighboring vesicle (Fig.~S5). While the number of exchanged lipids was not enough to observe significant mechanical perturbations on the timescales sampled in our simulations ($\sim$1 µs), over the course of seconds to minutes the degree of exchange would be extensive. Even at the timescale sampled, the rate of lipid ejection increased throughout the trajectory (Fig.~S5 C), suggesting this effect would be more significant on the millisecond timescale and beyond.

Alternatively, stalk initiation may depend on external energy input leading to a local depletion of ions or solvent at the adhesive interface between vesicles, which led to the rapid initiation of a stalk in our simulations here. Evidence for the latter mechanism comes from experiments where addition of EDTA led to rapid fusion of vesicles that were first allowed to adhere in the presence of Ca$^{2+}$~\cite{holz_effects_1979}. 

Future simulations and experiments with small amounts of amphiphiles containing two acyl chains, such as diacylglycerol or phospholipids, could determine the effect that splay could have on the fusion of fatty acid vesicles. Additionally, investigating alternative means of inducing disorder in the vesicle-vesicle interface, such as increased temperature, shear, or depletion of ions, could reveal alternative routes of fusion in fatty acid vesicles.

\subsubsection{Role of membrane mechanics}
Another focus of future work should be the impact of the mechanical properties of fatty acid membranes on adhesion. The chain length of fatty acids in vesicles can affect the propensity for fusion via the vesicle dynamics and interactions. The enhanced deformability associated with the short-chained vesicle in the simulations between a short-chained vesicle and long-chained vesicle resulted in a more extended adhesive interface, while the presence of micelles in simulations not only increased the interfacial area but facilitated near-instantaneous contact between vesicles. On a population scale, these effects are likely to increase the chance of a stochastic fusion initiation event.

Given the morphology of the interface varied between Sys1, Sys2, and Sys3 even in the same salt concentration and temperature, it is likely that the difference in mechanical properties between short-chained and long-chained vesicles would result in significant deviations in interfacial morphology upon changes in environmental conditions. Additionally, the formation of an adhesive interface has been shown to alter the membrane tension and thus diffusive properties of membranes, where an increase in diffusion rate was observed upon adhesion~\cite{oda_liposomal_2020}. Since we saw a correlation between diffusion and interfacial composition in our simulations, these differential changes in mechanical properties upon adhesion likely have implications for the dynamics of a hypothetical population of protocells. Performing experiments and simulations in various salt concentrations and temperatures, as previously done for phospholipid vesicles~\cite{murakami_morphologies_2020}, could further probe the relationship between membrane composition and adhesive properties in fatty acid vesicles of different chain lengths.  

 

\subsection{Hemifusion stability}
Curiously, even when we could adhere fatty acid vesicles to each other using typical fusogens such as divalent cations or depletants, full fusion and content exchange did not occur. The destabilization of the hemifused intermediate has been shown to be an essential step in forming a full fusion pore~\cite{marrink_mechanism_2003, knecht_molecular_2007}. Our simulations reveal that the formation of this hemifused intermediate occurs over similar timescales for long-chained and short-chained fatty acid systems, but only the system containing short-chained fatty acids exhibited a destabilization of the hemifusion diaphragm and formation of a full fusion pore after $\sim$50 ns, while the long-chained system remained in the hemifused state. Our experimental results also demonstrate the unusual stability of the hemifused state for pure oleic acid systems. 

Marrink and Mark demonstrated that the stability of the hemifusion diaphragm in phospholipid vesicles depended heavily on the degree of heterogeneity in the membrane composition, where the hemifused state remained stable for at least 100 ns for pure POPE vesicles, while in a mixed POPE/DPPC system it destabilized to a fusion pore in 15 ns~\cite{marrink_mechanism_2003}. Membrane heterogeneity appears to be a general factor in the stability of this intermediate state and thus in the dynamics of fusion. Further simulations and experiments could illuminate the role that heterogeneity plays in the stability of the hemifusion diaphragm in monoacyl lipid systems, where flip-flop can readily relieve bending stresses, unlike in pure phospholipid membranes~\cite{bruckner_flip-flop-induced_2009}. How more diverse chain lengths, head groups, degrees of unsaturation, or varying environmental conditions affect these processes will be the focus of future work.

\subsection{Fusion outcomes and timescales}
We attempted to capture fusion between vesicles using overnight timelapses (Fig.~S6, Movie~S1), and found that while vesicles that appear touching can remain stable for hours, fusion potentially happens within 30 seconds. In future, confocal video microscopy can be used to experimentally measure the timescales of hemifusion~\cite{nikolaus_direct_2010}. Because fusion is a rare event in these fatty acid systems even under the most fusogenic conditions, the use of micropipettes to force vesicle contact or live fusion-detection may be necessary to avoid high frame rate data acquisition over extremely long time periods.

We also characterized the sizes of vesicles before and after fusion, and found that the size distribution of vesicles is widespread. Although the average size seems to increase slightly after fusion, indicating an increase in volume as expected, the difference may not be statistically significant (Fig.~S7). This result indicates that there may not be a strong preference for what sizes of vesicles are prone to fusion. Given that all observed vesicles are in the giant regime, we would hypothesize any curvature stresses to be more significant for nanoscale vesicles, where size-dependent fusion may be more likely to occur. Investigating fusion phenomena for nanoscale vesicles may prove a fruitful direction for future work.

Together, our simulations provide a molecular picture of a vesicle system that is more poised for fusion due to enhanced probability of interactions and disruption of membrane packing in a short-chain system, while our experiments identify the source of this disruption to be lipid exchange resulting in an increase in fusion products. Furthermore, our simulations show that a local hemifused intermediate leads to rapid formation of a fusion pore between a short-chained and long-chained vesicle but not two long-chained vesicles, correlating well with experimental work that shows the highest fusion rates when two populations of vesicles made from different lipids mix.

\section{Conclusion}

We found that a range of typical fusogens, centrifugation, and agitation are not enough to promote fusion between fatty acid vesicles. Moreover, we found that hemifused intermediates are very stable for fatty acid vesicles, persisting for months. These properties reflect a major difference between fatty acid and phospholipid vesicles, possibly linked to the inability of fatty acids to undergo lipid splay, a critical step for phospholipid membrane fusion. Instead, we found that the most effective method for full fusion of fatty acid vesicles, leading to content exchange, was to add vesicles made from a different fatty acid. Simulations revealed that lipid chain length can impact membrane diffusion and mechanics, with the presence of shorter lipids decreasing the stability of a hemifusion diaphragm leading to the formation of a fusion pore. Adding fatty acid micelles also encouraged vesicle fusion, if samples were also agitated overnight, correlating well with simulations showing that excess surface area leads to enhanced contact between vesicles, which can then increase the probability of stalk initiation. The overall method for fusing model protocells containing fatty acids may therefore be one that disturbs lipid packing: enough to lower the energy barrier for fusion, but not so large as to lose vesicle contents to the environment.


\section{Experimental}
Given concentrations refer to the concentration of lipid, not the concentration of vesicles. \\

\textbf{Perturbing OA vesicles by adding potential fusogens}\\
Vesicles were made using the method described by Kindt and coworkers~\cite{kindt_bulk_2020}. In brief, 5 mM oleic acid vesicles were prepared in 50 mM Na-bicine buffer (pH 8.3) and 100 mM sucrose, encapsulating either 1 mM 8-hydroxypyrene-1,3,6-trisulfonic acid trisodium salt (HPTS or pyranine) or 2 mM calcein blue. They were then diluted ten-fold into a buffer containing 50 mM Na-bicine buffer (pH 8.3) and 100 mM glucose, prior to mixing the two populations together. The mixture was then exposed to the different conditions outlined in Table S1 by adding small quantities of concentrated stock solutions of each fusogen.

\textbf{Preparing palmitoleic acid (PA) vesicles}\\
For palmitoleic acid (PA) GUVs, 7.3 µL of PA at room temperature was added to a solution of sodium hydroxide (5.0 µL, 5 M) and Milli-Q water (244 µL) to form palmitoleic acid micelles (0.10 M). 

200 µL of the palmitoleic acid micelles (0.10 M) were then added to a solution of 1 M Na-bicine buffer (100 µL, pH 7.89) and Milli-Q water (690 µL) marked by 8-Hydroxypyrene-1,3,6-trisulfonic acid trisodium salt (HPTS) (10 µL, 0.1 M). The sample was put on the orbital shaker at 100 rpm (PSU-10i Grant Bio, UK) overnight to form palmitoleic acid vesicles (20 mM).

The 100 mM palmitoleic acid vesicle solution was prepared by adding 14.6 µL of PA to 1 mL of 2 mM calcein blue 100 mM Na-bicine (pH 8.3) solution, then vortexing for 10 s. 1 µL was then deposited gently into 150 µL of a 0.5 mM OA GUV sample encapsulating 1 mM HPTS, then imaged.

\textbf{Preparing oleic acid (OA) vesicles}\\
Oleic acid (OA) (15.9 µL) at room temperature was added to 12.5 µL of 5 M NaOH and Milli-Q water (488 µL) to form oleic acid micelles (0.10 M). 

200 µL of the oleic acid micelles (0.10 M) were added to a solution of 1 M Na-bicine buffer (100 µL, pH 7.89) and Milli-Q water (690 µL) marked by HPTS (10 µL, 0.1 M) or xylenol orange (XO; 10 µL, 0.1 M) respectively. The sample was put on the shaker overnight to form oleic acid vesicles (20 mM).

\textbf{Preparing membrane-dye-labeled oleic acid (OA) vesicles}\\
The paper rehydration method~\cite{kresse_novel_2016, lowe_methods_2022} was used to create 5 mM oleic acid vesicles in 50 mM Na-bicine (pH 8.4) containing either 20 µM TopFluor PC or Liss Rhod PE as membrane labels. 

\textbf{Preparing oleic acid (OA) micelles}\\
Oleic acid (OA) (7.9 µL) at room temperature was added in the solution of sodium hydroxide (6.25 µL, 5 M) and Milli-Q water (244 µL) to form oleic acid micelles (0.10 M). 

\textbf{Preparing `dummy' micelles}\\
Sodium hydroxide (6.25 µL, 5 M) was mixed with Milli-Q water (251.9 µL) to create `dummy' micelles.

\textbf{Fusing PA vesicles and OA vesicles} \\
Procedure: OA-XO vesicles (25 µL, 20 mM) and PA-HPTS vesicles (25 µL, 20 mM) were added to the solution of Na-bicine buffer (95 µL, 1 M) and Milli-Q water (855 µL). The solutions were equilibrated for 30 minutes and then observed under the microscope.

\textbf{Perturbing OA vesicles by adding oleate micelles or `dummy' micelles}\\
5 mM perturbation: 25 µL of 20 mM OA vesicles marked by HPTS (green) and 25 µL of 20 mM OA vesicles marked by XO were added to a solution of Na-bicine buffer (99 µL, 1 M) and Milli-Q water (805 µL). The mixtures were put on the orbital shaker at 100 rpm. After leaving it overnight, either oleate micelles (50 µL, 0.1 M) or dummy micelles (50 µL) were added to the mixture.

10 mM perturbation: 25 µL of 20 mM OA vesicles marked by HPTS (green) and 25 µL of 20 mM OA vesicles marked by XO were added to a solution of Na-bicine buffer (95 µL, 1 M) and Milli-Q water (755 µL). The mixtures were put on the orbital shaker at 100 rpm. After leaving it overnight, either oleate micelles (100 µL, 0.1 M) or dummy micelles (100 µL) were added to the mixture.

These samples were then put on the shaker (100 rpm) for 1 hour. After the sample was kept still on a bench for 10 minutes, an aliquot of the sample was then observed under the microscope. These samples were then put back onto the shaker overnight. After the sample was kept still on a bench for 10 minutes, an aliquot of the sample was then observed under the microscope. All images were analyzed with FIJI~\cite{schindelin_fiji_2012}. 

For further details please see the SI.

\section{Methods: Molecular Dynamics Simulations}

All vesicles were built using packmol~\cite{Martinez2009-om}, with a tolerance of 2.0 and discale of 1.4. For all long chain vesicles, fatty acid molecules were modeled as four hydrophobic beads (C1 Martini bead type), representing the acyl tail, and either a polar (P4 Martini bead type) or negatively charged head bead (Qa Martini bead type). The initial long chain vesicle was constructed by placing 3,605 negatively charged molecules and 3,605 neutral molecules in the inner leaflet, and 5,170 negative and 5,170 neutral molecules in the outer leaflet. Molecules were placed by constraining the final tail bead to a radius of 156 Å and the head bead to 139 Å for the inner leaflet, and the final bead to 160 Å and the head bead to 177 Å for the outer leaflet. For the initial short-chained vesicle, fatty acid molecules were modeled as three hydrophobic beads, and either a polar or negatively charged head bead. Short-chained vesicles were constructed by placing 4,200 negative and 4,200 neutral molecules in the inner leaflet, and 5,500 negative and 5,500 neutral molecules in the outer leaflet. Molecules were placed by constraining the final tail bead to a radius of 162 Å and the head bead to 150 Å for the inner leaflet, and the final bead to 165 Å and the head bead to 177 Å in the outer leaflet. The final density of molecules was chosen for each type of vesicle such that vesicles of the same radius did not exhibit large pores in the membrane upon short ($\sim$2 ns) equilibrations.

After initial structures of the short-chained and long-chained vesicle were generated, they were separately placed in a waterbox with 10 Å padding in each dimension and ionized with 100 mM NaCl. Water beads were represented with P4 Martini bead types, and antifreeze particles were added as BP4 bead types. Na$^+$ ions were represented with QD Martini bead types, and Cl$^-$ as QA bead types. Both vesicles were minimized for 1000 steps and equilibrated for $\sim$2 ns. At the end of the equilibration, the vesicles were saved as separate structures. From these equilibrated vesicles, the final systems were constructed. A long chain vesicle (vesicle 2) was placed 370 Å in the +\textit{x} direction from the center of mass of the short chain vesicle (vesicle 1) and both were saved as a single structure (short chain/long chain system; Sys1). Similarly, a long chain vesicle (vesicle 2) was placed 370 Å in the +\textit{x} direction from the center of mass of a long chain vesicle (vesicle 1) and both were saved as a single structure (long chain/long chain system; Sys2). These two systems were solvated with 48 Å padding in the \textit{y} and \textit{z} axes (orthogonal to the axis that runs through both vesicles), and 30 Å in +\textit{x} and -\textit{x}. For the system with micelles, an in house Tcl script was written to add individual molecules to the short chain/long chain system by aligning to water beads, with any clashing water molecules removed (Sys3 short/long/micelles). All three of these systems were neutralized and ionized with 100 mM NaCl.

All simulations were performed using NAMD 2.14~\cite{Phillips2005-oe} with the Martini force field~\cite{Marrink2007-hi}. Analysis of simulations and generation of molecular figures was done using visual molecular dynamics (VMD)~\cite{Humphrey1996-wo}. All simulations were performed in the NpT ensemble, using a 20 fs timestep and a cutoff for van der Waals and electrostatic interactions set at 12 Å. All three systems were first minimized for 1000 steps and equilibrated for 50 ns. Sys1 (short/long) and Sys3 (short/long/micelles) were additionally equilibrated to a total of 500 ns. After the initial 50 ns equilibration, constant-velocity steered molecular dynamics (SMD) was applied in the +\textit{x} direction to the center of mass of vesicle 1 in Sys1 (short/long) and Sys2 (long/long) for $\sim$125 ns, using virtual springs with a stiffness of \textit{k}$_{s}$ = 1 kcal mol$^{-1}$ Å$^{-2}$ and a pulling speed of 0.1 nm/ns. After SMD, when a completely adhered state was achieved for both Sys1 (short/long) and Sys2 (long/long), both systems were freely equilibrated for an additional 956 ns. 

Initiation of a stalk was induced in Sys1 and Sys2 by applying an outwardly-directed radially symmetric force to the solvent and ion beads directly in the interface between vesicles starting at the end of the 956 ns equilibration. This force was applied using NAMD Tcl interface, and only those water or ion beads that were found within a circle of radius 20 Å of the center of the interface was pushed outward at any step of the simulation. This force was maintained throughout the subsequent 51 ns.

For summary of simulation conditions please see Table S2.

\medskip
\textbf{Supporting Information} \par 
Supporting Information is available from the Wiley Online Library and includes supporting figures, a table, and methods.

\medskip
\textbf{Acknowledgements} \par 
A.W. is supported by an Australian Research Council Discovery Early Career Award DE210100291. J.W.S. is an Investigator of the Howard Hughes Medical Institute.  C.N. is an HHMI Research Associate. All simulations were performed using the Midway2 and Midway3 supercomputers at the Research and Computing Center at the University of Chicago.

\medskip




\section{References}
\bibliography{manuscript}

\end{document}